# SOCIO-MATERIAL NETWORK ANALYSIS:

## A Mixed Method Study of Five European Artistic Collectives

**Nikita Basov**

**St. Petersburg State University**

*Abstract*

In this paper, I argue that we can better understand the relationship between social structure and materiality by combining qualitative analysis of practices in shared physical space with statistical analysis. Drawing on the two-mode approach, I treat social and material structures together with the relationship between them as a two-level socio-material network. In a mixed method study, formalized ethnographic data on such networks in five European artistic collectives are subjected to multilevel exponential random graph modelling. It sheds light on how different types of interpersonal ties condition the engagement of individuals with similar materiality over time.

*Keywords:* Socio-material network, physical object, object usage, material structure, two-mode approach, exponential random graph model, mixed method

# 1. Introduction

Sociologists are well aware of the role materiality plays in the reproduction of the social order. By creating and using physical objects, social actors draw on the corporeality of the material to reinforce the social with its embodied durability, channel symbolic power, represent socio-cognitive constructs, communicate and reproduce standards for social interaction (Bourdieu, 1984, 1990, 1996; Luhmann, 2000; Mead, 1912). However, little is known about how interpersonal ties affect engagement of individuals with shared materiality in their everyday practice. Understanding this role could shed light on the relationship between social structure and materiality at the micro level.

Up until now, network analysis that traditionally examines concrete structures of social ties, paradoxically, has not considered how these affect the ways individuals utilize physical objects, which are no less concrete. Qualitative micro-perspectives have covered more ground in empirical investigation of materiality (Berns, 2016; Craig, 2011; Griswold et al., 2013; Jarness, 2015; Maisonneuve, 2001; Newton-Francis and Young, 2015). Using ethnographic data they argued that the relations between material things comprise the texture of materiality just like inter-personal ties constitute the texture of the social and hence should not be neglected, and even that materiality has its own structuring logic (Callon et al., 1986; Latour, 2005). For example, Maisonneuve (2001) analyzes how in the 1920-30s, gramophones, records and catalogues connected by new practices of collecting and listening changed the appreciation of music and created new markets.

To address the relationship between two self-consistent orders, such as social network ties and the structure of materiality, the lens of duality perspective (Breiger, 1974; Simmel, 1955) is useful. Applied in various contexts (Breiger and Pattison, 1986; Martin, 2000; Mohr, 2000; Mohr and Duquenne, 1997; Mohr and Neely, 2009), 'dual' thinking considers pairs of orders jointly, as co-constituting but also self-organizing. The most broadly known application of the duality thinking is, perhaps, the culture and structure duality (Breiger, 2000; Martin, 2000; Mohr and Neely, 2009; Mohr and Rawlings, 2010; Schultz and Breiger, 2010), that "involves recognizing that neither culture nor social structure, system of meaning nor mode of practice, should be privileged in the analysis. Rather these multiple orders of social life need to be given equal weight and each should be seen as being constituted by its embeddedness within the other." (Mohr, 2000).

Given that the distinction between the social and the physical orders is no less explicit, but at the same time the two orders are known as interrelated, it makes perfect sense to extend dual thinking to study the relations between materiality and social structure. Schweizer (1993) has already applied the dual logic to study the interplay between materiality and social structure in urban communities of French Polynesia, hunter–gatherers in Zaire, and peasants in rural Java. By examining the dual relationship between material possessions owned by community members and their social ranking, he found that the ordering of one was interdependent with the ordering of the other. More recently, Mohr and White (2008) drew on Schweizer's work to examine the institutional stability of Indian caste systems and found this stability to be based on the dual ordering of social networks and cultural values realized through transactions involving food, everyday items, water, garbage, etc.

Methodologically, the dual perspective is most frequently operationalized using the two-mode approach, introduced by Breiger (1974), and its generalization (Fararo and Doreian, 1984). Successful in the studies of dualities (Breiger, 2000; Martin, 2000; Mohr and Duquenne, 1997; Mohr and Neely, 2009), and already applied to link individuals and objects (Mohr and White, 2008; Schweizer, 1993), this approach can be used to represent the duality of the social and the material. Namely, the two-mode network of object usage connecting actors to objects captures the relationship between the two orders, represented by one-mode networks: showing social ties on the one hand and links between physical objects on the other. We can then analyze 'socio-material networks' (see Figure 1) that include the three types of relations: (1) social ties, (2) links between material objects as they are combined and collocated in the physical space, thus comprising the structure of materiality, and (3) links between actors and objects they use throughout their material activities.

Such an extension allows a formal representation of specific network patterns showing how socially connected actors utilize objects and, this way, jointly engage with material structure. For example, when two collaborators use certain tools to accomplish their work tasks we can view it as a two-level cycle, where nodes are two objects and two actors, two edges are object usages, one edge is a collaboration tie and one edge represents an association between tools normally used in the course of the work. Examining the relative importance of such patterns enables insight into the specific principles, according to which social ties affect joint engagement of individuals with material structure as they use physical objects in everyday practice.

This can be done statistically—if we (technically[1]) approach socio-material networks as two-level networks, hence treating the social and the material as distinct orders interconnected by the two mode object usage links. Then, it is possible to apply such a state-of-the-art technique of network analysis as multilevel exponential random graph modelling (MERGM) (Wang, 2013; Wang et al., 2013). MERGMs enable addressing dually co-constitutive orders with the statistical power that was not available before and inquire about particular principles of how social ties affect engagement with shared materiality.

---

[1] Note, however, that similarly to Brennecke and Rank (2017) while using the opportunities provided by MERGMs, I do not think of the social and the material as nested levels.

Because the relation between social structure and materiality is not yet transparent, we propose an extension of the mixed method approach to data collection and analysis, oscillating between the qualitative and the quantitative (see Fuhse and Mützel, 2011; Godart and Mears, 2009). It starts with gathering a mixed set of data that includes both social network survey data and ethnographic data, such as interviews and visual observations. This mixed data is used to produce the two-level networks subjected to statistical modeling. And then, the modeling results are contextualized using the ethnographic data.

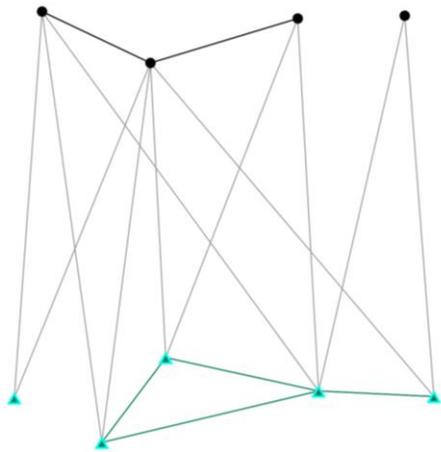

**FIGURE 1. A SOCIO-MATERIAL NETWORK**[2]

**Note:** Circles: actors. Triangles: objects. Black lines: social ties. Cyan lines: material links. Grey lines: usages of objects.

The present paper aims to test the outlined approach and thus open up a research avenue rather than to draw conclusions. With this in mind, I focus my empirical inquiry on artistic collectives. Creatives have been the subject of network analysis before (Basov et al., 2016; Comunian, 2011; Crossley, 2009; De Nooy, 1999, 2002; McAndrew and Everett, 2015). Although the relation between the social and the material is clearly not limited to small groups of visual artists sharing spaces which I analyze, such a setting allows a close ethnographic examination of how social structure molds materiality within a reasonable amount of resources. Firstly, the role of space and materiality for artists, despite the technological changes of the 20th century in communication and transportation, is widely recognized (Carlozzi et al., 1995; Griswold et al., 2013; Oberlin and Gieryn, 2015; Peter, 2009) and we can expect artists to be explicit in their material activities. Secondly, in the artistic setting, the social and the material are not constrained much by the formal regulations, and hence are less dependent on organizational contexts.

---

[2] ORA NetScenes was used to produce all network plots (Carley at al., 2013).

To induce generalizability of the findings, I use the data on five collectives of artists located in different European cities – Barcelona, Hamburg, London, Madrid, and St. Petersburg, which are embedded in different social and cultural contexts. Still, I realize that the given setting does not allow for broad conclusions to be made about the overall relationship between social and material structure.

The paper proceeds as follows. I start by presenting hypotheses on the basic principles of socio-material structuring, suitable for the purposes of illustrating the proposed approach. Then, I describe the empirical setting and the techniques of data collection and analysis. After that, results of the analysis are presented. I conclude with discussion of the results and an outlook on limitations and future prospects.

## 2. Principles of Socio-Material Structuring

While the effect of materiality on interpersonal relations has been widely argued and empirically investigated by the qualitative approaches (Berns, 2016; Callon et al., 1986; Craig, 2011; Griswold et al., 2013; Latour, 2005), the seemingly obvious inverse dependency—the ways in which social ties draw on common materiality—still needs empirical inquiries. Since the material provides the social with a 'body', we need to understand how this embodiment happens. Hence, my hypotheses revolve around the question of how social ties affect engagement of individuals with similar materiality in a shared space over time.

Informed by the previous studies, I distinguish between two types of social ties: collaborations and emotional attachments. Whereas these two kinds of relations are not mutually exclusive (Balkundi and Harrison, 2006), interactions specific for them clearly differ, implying specific goals, relational frames, and expectations. Collaborations rather correspond to joint work and exchange of work-relevant information and advice, while emotional attachments are more affect-based and involve joint leisure, moral support and mutual care (Balkundi and Harrison, 2006; Ibarra, 1995; Lincoln and Miller, 1979; Umphress et al., 2003). Hence, collaborations and emotional attachments are known to have different effects (Fang et al., 2015; Umphress et al., 2003). The ways materiality is engaged with in these two different relational contexts can also be expected to diverge.

Since little is known on the particular patterns of how social ties affect engagement with material structure via usage of objects in shared spaces, in this inquiry I draw my hypotheses against the background of network patterns and basic principles known to structure social networks, such as contagion and homophily.

It is now commonplace that behavior, tastes and preferences of actors depend on those of their peers in social networks (DiMaggio, 1987; DiMaggio and Useem, 1978; Erickson, 1996; Haynie, 2001; Marsden et al., 1982; Warr, 1996), especially when strong interpersonal ties are involved (Erickson, 1988; Haynie, 2001; Mark, 1998). This argument is largely based on a principle known under different names: relational proximity (Borgatti and Foster, 2003; Meyer, 1994; Rice, 1993), socialization (Kandel, 1978), transmission (Mark, 1998) and contagion (Krohn, 1986). For the sake of readability, I will further refer to it as 'contagion'. Simply put, contagion implies that interacting people tend to learn from each other about existing behavior possibilities (Mark, 1998) and replicate each other's choices, merely imitating their alters (Akers, 1985), being talked into making similar choices or conforming to the group (Cohen, 1978).

It may be the same for the choices of physical objects people make. For instance, we often order the same items in a bar or a restaurant just because our peers have told us about them or because we have observed their preferences, which makes our choice easier when we follow them or, in some cases, to avoid conflict with our peers, e.g., when everybody drinks whiskey one is less likely to go for a beer. Perhaps, a particularly strong contagion effect can be expected when shared materiality is involved, because the possibility of direct observation of each other's behavior stimulates imitation. It is empirically shown that the behavior of peers has a stronger influence on an ego's behavior than peers' attitudes (Warr and Stafford, 1991).

There is another ability that physical things have to induce chances for interpersonal ties to stimulate embeddedness in the same material things. Objects remain despite the passing of time. Therefore, individuals often utilize things to commemorate their relationship itself or certain aspects of it. For example, when two lovers, friends, or relatives spend time together, they draw on certain items to represent their association. For instance, collections of presents remind lovers of their happy moments, persistently evoking memories, and tend to be got rid of (returned or even destroyed) when the relationship is ended.

Studies have demonstrated that both work-related (Carley, 1986; Carley, 1991; Rice and Aydin, 1991; Umphress et al., 2003) and emotional (Kilduff, 1990; Krackhardt and Kilduff, 1990; Pastor et al., 2002; Umphress et al., 2003) ties between individuals can stimulate joint activities.

Combined, this argument brings me to the first pair of hypotheses:

**Hypothesis 1a:** Actors connected by collaboration ties are likely to use the same material objects in shared spaces.

**Hypothesis 1b:** Actors connected by emotional attachment ties are likely to use the same material objects in shared spaces.

Furthermore, following Lazarsfeld and Merton (1954), a large number of network-analytical studies use the notion of homophily to indicate a principle of network structure formation, inverse to the one known as contagion. Owing to homophily, individuals with similar stable characteristics, such as gender or education, are more likely to be connected by interpersonal network ties (Heider, 1958; Kandel, 1978; McPherson et al., 2001). The debate on casual sequencing relating homophily and contagion has a long history (see, e.g. Haynie, 2001; Kandel, 1978; Mark, 1998), and among others is posed with regard to dual relationships between cultural preferences and social network structure (Vaisey and Lizardo, 2010). In short, the question is whether individuals choose their alters because of similarity in characteristics (homophily), or whether social ties induce similarity of characteristics (contagion). Contagion and homophily have also been shown to operate simultaneously (Kandel, 1978; Mark, 1998; Vaisey and Lizardo, 2010). As engagement with materiality is not a stable characteristic, which makes homophily not applicable directly, and as here I am rather interested in the influence of social ties on materiality, the causality between contagion and homophily is beyond the scope of this paper. Yet, with regard to material objects it seems to be of interest to test for a hybrid principle (see Mark, 1998), i.e., the impact of homophilous ties (ties between individuals with similar stable characteristics, such as gender and education) on achievement of similar material preferences. For example, we can expect two collaborating artists of the same gender, who underwent similar training, and who work in the same genre to use similar sets of tools in their work.

Since homophily is known to operate in different types of network ties (for an overview see McPherson et al., 2001), the described logic is likely to work both with regard to collaborations and emotional attachment ties.

**Hypothesis 2a:** Actors with similar stable characteristics who are connected by collaboration ties are likely to use the same material objects in shared spaces.

**Hypothesis 2b:** Actors with similar stable characteristics who are connected by emotional attachment ties are likely to use the same material objects in shared spaces.

Finally, objects are rarely isolated in physical spaces. Dozens of items comprise physical contexts, often lying in proximity to each other and related functionally. For instance, teapots are linked to cups, pens – to paper, computers – to printers. Numerous relations between objects create the canvas of materiality, which is the physical scene for the realization of social relations. For single individuals and dyadic social ties this material context is largely a given: it is persistently present when individuals enter the space and when they leave, it is molded by others, it imposes constraints determining which combinations of objects are possible and which are not, and so forth.

Still, the social strives to gain material endurance. Individuals tend to 'settle' in different areas of the material landscape – both to achieve certain goals and unintentionally. Although no corresponding empirical tests have been found in the literature, we can assume that persons' engagement with material structure is affected by social structure at the level of dyadic interpersonal ties. When two individuals continuously interact and operate with objects they may be expected, purposefully or unintentionally, to get embedded in the materiality that suits their relationship. Associations between objects that the dyad engages with comprise structures of individuals' common material context, as throughout their interactions people use things that go together and/or are placed close to each other. For example, take two mosaic artists occupying the same workshop, sitting next to each other on their chairs gossiping about their fellow artists - one is likely to reach for a hammer to cut off mosaic pieces from the same ceramic pot that her pal takes a chopper from. Here, not only the object (the ceramic pot) is shared, but a number of other objects used (but not necessarily shared) get associated (hammer and pot, chopper and pot, two chairs, etc.) because people like to hang out together, although not collaborating on anything. This leads to the third pair of hypotheses:

> **Hypothesis 3a:** Actors connected by collaboration ties are likely to be engaged with the same material contexts in shared spaces.

> **Hypothesis 3b:** Actors connected by emotional attachment ties are likely to be engaged with the same material contexts in shared spaces.

The patterns corresponding to the principles of socio-material structuring captured by the hypotheses are summarized in Table 1.

**TABLE 1. PATTERNS OF SOCIO-MATERIAL STRUCTURING**

| PRINCIPLE | Influence of dyadic social ties on object sharing [H1] | Influence of homophilous social ties on object sharing [H2] | Influence of dyadic social ties on engagement with the same material context [H3] |
|---|---|---|---|
| ILLUSTRATION | (triangle with two actors sharing one object) | (two same-attribute actors sharing one object) | (two actors each engaged with separate objects in same context) |

**Note:** Circles: actors. Numbered circles: actors with same attributes. Triangles: objects.

It is crucial for the remainder of this paper that my hypotheses are not mutually exclusive, because the principles they imply may operate simultaneously, altogether yielding the overall socio-material structure. For instance, dyads may not tend to share objects overall (H1), while dyads of similar gender do (H2). Moreover, these principles may reinforce each other or compete.

## 3. Data and Method

### 3.1. Empirical setting and data collection

Empirical analysis of this paper draws on ethnographic studies of five artistic collectives located in Barcelona, Hamburg, London, Madrid and St. Petersburg (see also: Basov and Brennecke, 2017; Nenko, 2017; Nenko et al., 2017; Pivovarov and Nikiforova, 2016). Art groups are usually informal and changeable in structure, flexible in the establishment and severing of social ties. They also allow diverse usage and combining of objects. This makes a contrast to more formalized settings, such as organizations, that are the focus of most empirical studies accounting for materiality and are known to constrain personal decisions (Stigliani and Ravasi, 2012).

Furthermore, all of the collectives make visual artworks (painting, graphics, sculpture, mosaics, photography, video, installations, performances, and other) and live and/or work in shared spaces. It implies that members of the collectives have the material environments that they can potentially use to embed their social ties in, and that ethnographers, upon accessing the field, can observe these processes.

In each of the cases, the field researchers discovered spaces ranging from dozens to hundreds of square meters and filled with thousands of material objects. As expected, these objects were both work-related, such as artworks, tools, equipment and artistic materials, and everyday items, like books, papers, journals, household objects, furniture, dishes, food, clothes, and consumer electronics (see Figure 2 for some illustrations).

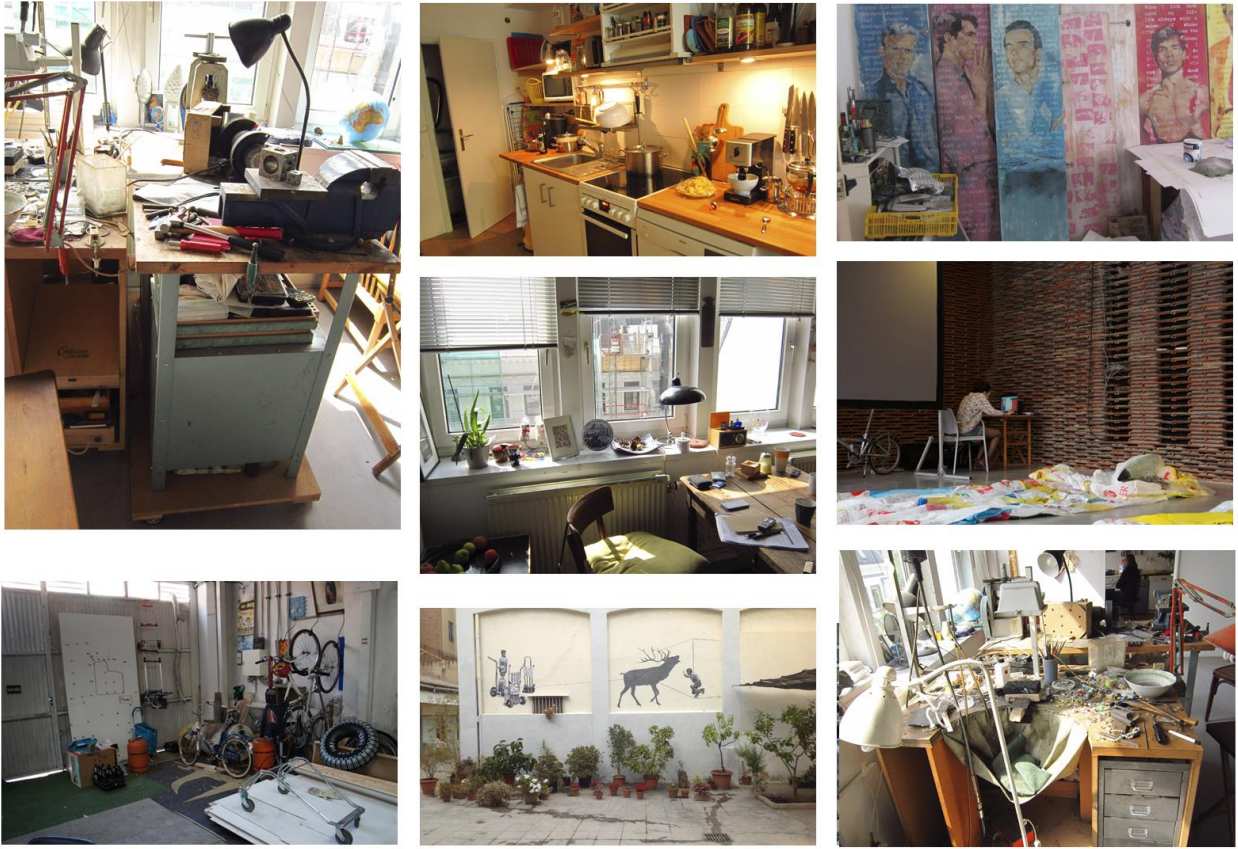

**FIGURE 2. OBJECTS IN THE ARTISTIC SPACES OF BARCELONA, HAMBURG, LONDON, MADRID AND ST. PETERSBURG**

During data collection, the researchers observed how the artists engaged in discussions and joint projects, exchanged information, casually interacted and hung out with friends, witnessing how they utilize, transform and exchange physical objects in these social processes (see Figure 3 for illustrations). This enabled an inquiry into how everyday and work-related interactions structure material contexts.

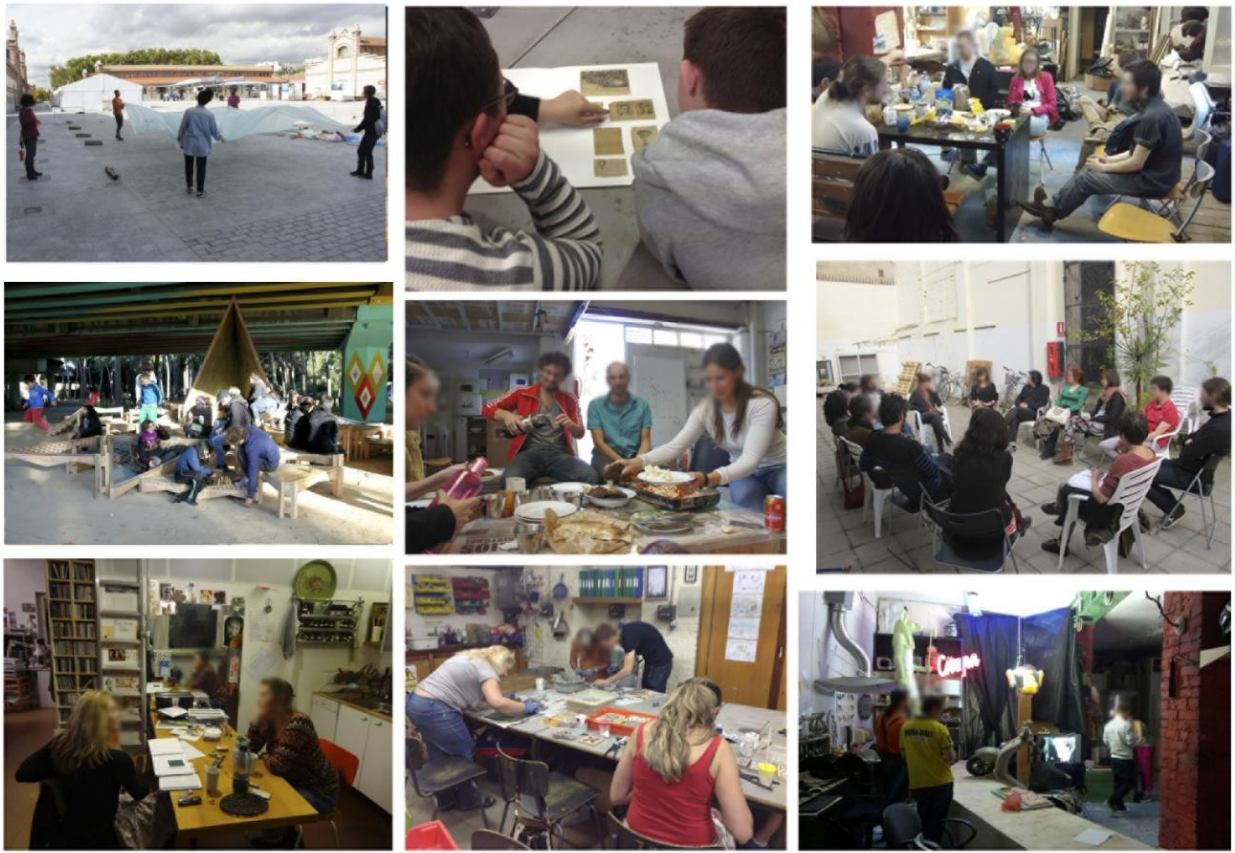

FIGURE 3. INTERACTIONS IN THE ARTISTIC SPACES OF BARCELONA, HAMBURG, LONDON, MADRID AND ST. PETERSBURG

The data I analyze in this paper was gathered in two waves with a six-month interval between them, the first wave being in Autumn 2014 and the second wave in Spring 2015. In each wave, the data was collected by five teams of professional sociologists in each of the five collectives simultaneously, following uniform procedures described below. Each wave lasted about three weeks.

The boundaries of the five collectives are quite flexible. While each group has a stable core of permanent members, many join for only for several months. Some leave after several years. The dataset includes only core members: those who have stable membership and continuous involvement.

In each wave, each of the studies started with an excursion around the space run by some of the members in each collective, providing first-hand information on the arrangement of the physical setting and members' material behavior in it. These were followed by semi-structured interviews (amounting to 125 in the two waves) with each member – to learn about their educational trajectories, the artistic genres they work in and the ways they use common spaces and objects. A separate (structured) section of the interviews addressed collaborations and emotional attachments between members, asking about their social ties to other members, as well as usages of objects and the ways objects are combined – both by themselves and others in the group.

Later on, the researchers conducted ethnographic observations of moderate participation (108 in two waves) of how creatives interact in shared material settings and conduct material activities there. Observation guidelines included detailed descriptions of objects, objects usage, references to objects in interaction, and location of objects in different functional zones.

During each wave, the research teams also conducted photo elicitations (113 in total) with each member, sequentially showing about 50 photos of the shared zones filled with objects to stimulate reporting about objects and their usage.

Finally, verified sociometric surveys using the roster method were run to capture social network ties in a particular wave. All the core members participated in the survey, which resulted in a 100% response rate.

Two types of social ties were inferred: collaborations and emotional attachments. The former involve joint work. Unlike them, the latter correspond to expressive relationships, such as friendships. For collaborations, we asked each person to mark everyone they had worked with at least once a week during the previous six months. For expressive ties, members were to indicate everyone towards whom they felt strong emotional attachment.

### 3.2. Data formalization and construction of the networks

Matrices corresponding to links between objects as well as between actors and objects were produced for each wave separately using the corresponding ethnographic data. When an object usage by one of the core members was directly observed or when an informant mentioned that he or she used an object during the preceding six months, a link between the actor and the object was registered. Each link was supplemented with data on regularity, duration and way of usage, when available.

Out of those, only objects used by at least two core members were included in the dataset.

Based on Riggins (1990), links between objects indicating their involvement in the same activities were based on functional relations between the items (i.e., being used together, e.g. a bottle and a glass) or on their physical proximity (continuous collocation, e.g., a painting near a sculpture). Similarly to object usage, the data sources were observations, photo elicitations and interviews.

The data on links was triangulated. For example, if during a photo elicitation one person mentioned that another person used an item, while in her own elicitation the latter person denied usage of such an object, then the corresponding link was not included in the dataset.

To get social network matrices, results of the verified sociometric surveys were triangulated as well, based on ethnographic observations and interview answers about ties with other members collected during the field study.

Data processing resulted in ten two-level socio-material networks (two networks corresponding to the two waves for each of the five collectives) that included (1) social ties of collaboration and friendship, (2) links between objects and (3) object usage.

## 3.3. Analysis

To test the hypotheses on the principles of socio-material structuring, I conduct statistical modelling using MERGMs: exponential random graph models for multilevel networks (Wang et al., 2013). In general, ERGMs express the probability of observing a specific network, using parameter estimates representing different network patterns, which correspond to certain principles of network structuring. Including multiple patterns allows mutual reinforcement and competition between different principles to be accounted for (Lusher et al., 2012; Zappa and Robins, 2016).

MERGMs are a class of ERGMs designed to model two unipartite networks and a bipartite network between them. They account for structuring in each of the three networks, simultaneously checking for dependencies involving several types of links.

Since I seek to understand how social structure influences engagement with the material, I estimate the models for artists' social ties in wave 1 and usage of objects and material structures in wave 2. Thus, reverse causality is excluded (for a similar approach, see Brennecke and Rank, 2017; Kim et al., 2016). Similarly, gender, education, and genre attributes are stable characteristics, acquired long before we documented usage of objects and material structures in the physical spaces of the collectives.

The nodesets are conformed across the two waves so that only actors and objects present in both waves and wave-specific relations between them are subjected to the analysis.

Furthermore, I aggregate networks of all five collectives to capture socio-material structuring principles invariant for the five groups (see also Basov and Brennecke, 2017). To take into account that the five groups are separate socio-material networks and hence only ties within the groups are to be accounted for, I used 'structural zeros' (Kalish and Luria, 2013) and thus excluded ties between the collectives from the estimation.

Two separate models are used to estimate emotional attachments and collaborations as predictors of the object usage and material context networks. Each of the models simultaneously includes patterns comprised of one type of relations and patterns combining several types of relations at the same time. They involve (1) the structure of the unipartite social network, (2) the structure of the unipartite material context network, (3) the structure of the bipartite object usage network and (4) cross-level dependencies of the three networks and exogenous effects of actors' attributes. It allows for testing the relative importance of different principles of socio-material structuring outlined above as they potentially reinforce or compete with each other, while accounting for various structuring processes within the social, material and object usage networks themselves. By including patterns that correspond to my three hypotheses, I check if the observed material structure forms because actors tend to share objects when they are connected with social ties (H1), and/or because being both connected with a social tie and sharing an attribute stimulates object sharing (H2), and/or because people tend to engage with common material contexts (H3), while controlling for structuring processes in all of the three networks, plus a number of other cross-level patterns. Table 2 describes the patterns included in the models.

Including the patterns proposed by Wang et al. (2013), I use MPNet (Wang et al., 2014) to estimate their relative contribution to link creation using the Markov-Chain Monte-Carlo maximum-likelihood method (Snijders, 2002) to produce parameter values for each of the patterns while conditioning their occurrence on the likelihood of observing the overall socio-material network.

To test the goodness of fit (GOF) of the models, I run conventional procedures (Hunter et al., 2008). Results of the GOF tests (see Appendix A) show that the models capture structure formation in the given networks quite well.

.

**TABLE 2. PATTERNS INCLUDED IN MERGMS**

| Illustration | Interpretation |
|---|---|
| 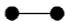 | Edge |
| 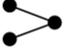 | 2-star |
| 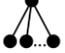 | Degree spread |
| 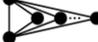 | Triadic closure |
| 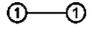 | Tie between actors with same attribute value |
| 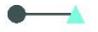 | Actor using object |
| 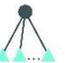 | Object usage degree of actors |
| 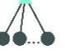 | Usage degree of objects |
| 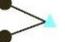 | Pair of actors sharing an object |
| 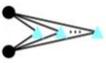 | Pair of actors sharing objects |
| 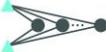 | Pairs of objects used by actors |
| 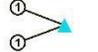 | Actors with same attribute values sharing object |
| 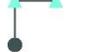 | Usage of objects that are part of material contexts |
| 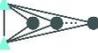 | Engagement with same materiality |
| 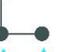 | Influence of social ties on usage of objects |
| 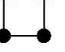 | Influence of dyadic social ties on usage of objects |
| 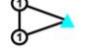 | *Influence of dyadic social ties on object sharing [H1]* |
| 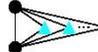 | *Influence of dyadic social ties between actors with same attribute values on object sharing [H2]* |
| 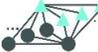 | *Influence of dyadic social ties on engagement with the same material context [H3]* |

**Note**: Black circles: actors. Numbered circles: actors with same attributes. Cyan triangles: objects. Circles' numbering indicates attributes of actors.

To account for multicollinearity issues with regard to those of the independent variables (patterns of social network ties with and without actor attributes) that did not show significant results when included in the same models,[3] I check correlation coefficients between estimates for these variables (see Appendix B). All of the correlations are small.

# 4. Results

This section starts with descriptive statistics. Then I present parameter estimates of the MERGMs, followed by discussion of the results, which starts with an overview of the patterns related to the three basic networks and proceeds to a more detailed consideration of the cross-level patterns involving combinations of the basic relationships and testing my hypotheses. Finally, the patterns are illustrated and reflected upon using the ethnographic data.

## *4.1. Descriptive statistics*

Table 3 contains descriptive statistics for the corresponding socio-material networks, including a summary of group members' attributes. Figure 4 and Figure 5 plot the social and the material orders in these networks, respectively.

On average, the material network in the collectives consists of 19 objects, the largest being 42. The average number of artists per collective is 8, reaching a maximum of 13. In some cases, few artists are present in both of the waves because the composition of the collectives significantly changed between the waves. Friendship and collaboration networks are comparable, both having quite high density, which is not surprising given that the collectives are small groups. Most of the social networks exhibit moderate centralization, amounting to 35-36 percent on average. The average degree, that is the number of ties individuals have, is 3.3 for the emotional attachments network and 3.15 for the collaboration network. The densities of the material networks are consistently low – mostly below 10 percent. Material network degree centralization is moderate, so no objects are engaged in significantly more material contexts than the rest. On average, an object is engaged in a relationship with one other object. As for the object usage network, the densities there are quite high and vary between 53 and 69 percent. Individuals are active in using objects. Each object is used by roughly 5 individuals on average and individual artists use between 6 and 23 objects.

---

[3] Note that in all linear and log-linear models multicollinearity is not an issue when variables show significant results when included in the same models.

TABLE 3. DESCRIPTIVE STATISTICS ON THE NETWORKS OF THE ARTISTIC COLLECTIVES

|  |  | Average | Min | Max | Total |
|---|---|---|---|---|---|
|  | Shared objects per collective | 19 | 9 | 42 | 94 |
|  | Individuals per collective | 8 | 4 | 13 | 39 |
| Emotional attachment | Density | 50% | 28% | 83% |  |
|  | Degree centralization | 36% | 29% | 48% |  |
|  | Average degree | 3.30 | 1.20 | 7.08 |  |
| Collaboration | Density | 56% | 27% | 100% |  |
|  | Degree centralization | 35% | 0% | 51% |  |
|  | Average degree | 3.15 | 2.00 | 3.79 |  |
| Material | Density | 7% | 3% | 12% |  |
|  | Degree centralization | 22% | 9% | 34% |  |
|  | Average degree | 1.00 | 0.60 | 1.43 |  |
| Object usage | Density | 61% | 53% | 69% |  |
|  | Average object degree | 4.59 | 2.56 | 6.98 |  |
|  | Average actor degree | 10.93 | 5.75 | 22.54 |  |
| Group member attributes | Female group members | 41% | 25% | 78% |  |
|  | Group members with artistic education | 85% | 69% | 100% |  |
|  | Diversity in genre* | 0.47 | 0 | 0.81 |  |

*Herfindahl index is reported, zero meaning that all artists with a group work in the same genre and one meaning they all work in different genres.

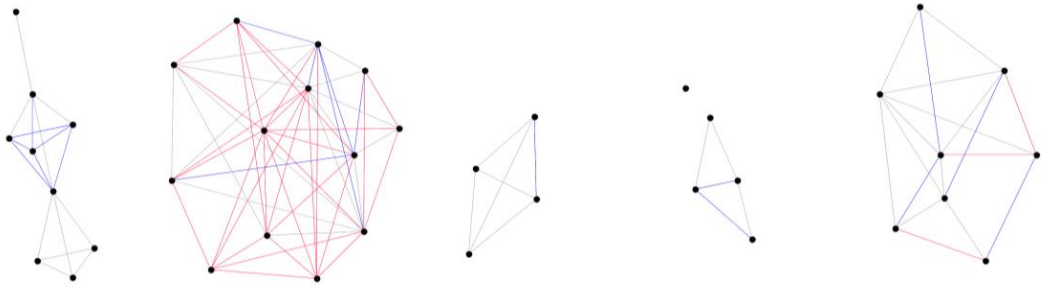

**FIGURE 4. SOCIAL NETWORKS OF THE ARTISTIC COLLECTIVES**

**Note:** Left to right: Barcelona, Hamburg, London, Madrid, St. Petersburg. Circles: actors. Blue lines: collaboration ties. Red lines: emotional ties. Grey lines: multiplex emotional and collaboration ties.

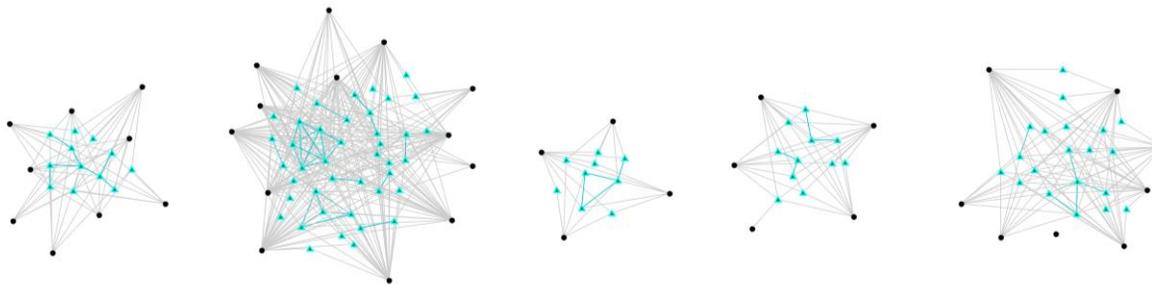

**FIGURE 5. MATERIAL NETWORKS OF THE ARTISTIC COLLECTIVES**

**Note:** Left to right: Barcelona, Hamburg, London, Madrid, St. Petersburg. Circles: actors. Triangles: objects. Cyan lines: material structures. Grey lines: usages of objects.

### *4.2. MERGMs results*

Table 4 displays the converged models.

Table 4. Results of MERGMs

| | Illustration | Interpretation | Collaborations | | Emotional attachments | |
|---|---|---|---|---|---|---|
| | | | Parameter | SD | Parameter | SD |
| Social | | Edge | 1.5175 | 1.463 | -4.6234*** | 1.692 |
| | | 2-stars | - | - | 0.2042** | 0.097 |
| | | Degree distribution | -2.1271*** | 0.505 | -0.4315 | 0.675 |
| | | Triadic closure | 1.5614*** | 0.385 | 0.6688* | 0.348 |
| | | Ties between actors of same gender | -0.0048 | 0.573 | 0.2982 | 0.635 |
| | | Ties between actors of similar education | -2.1758*** | 0.749 | 0.5965 | 0.951 |
| | | Ties between actors working in same artistic genre | -0.7751 | 0.725 | 2.6191*** | 0.998 |
| Material | | Edge | -5.1406*** | 1.127 | -8.1774*** | 1.393 |
| | | Degree distribution | -1.1035*** | 0.247 | -0.8538*** | 0.22 |
| | | Triadic closure | 1.9529*** | 0.27 | 2.0034*** | 0.268 |
| Object usage | | Actors using objects | -0.5113 | 1.161 | -4.6461*** | 1.199 |
| | | Object usage degree of actors | 1.2975*** | 0.439 | 3.0259*** | 0.702 |
| | | Usage degree of objects | -1.9827*** | 0.54 | -1.7115*** | 0.528 |
| | | Pair of actors sharing an object | 0.0447*** | 0.014 | - | - |
| | | Pair of actors sharing multiple objects | -0.0679* | 0.038 | - | - |
| | | Pairs of objects used by actors | - | - | 0.8427*** | 0.303 |
| | | Objects shared by actors of same gender | 0.113** | 0.045 | 0.2336*** | 0.062 |
| | | Objects shared by actors of similar education | 0.106** | 0.051 | -0.0269 | 0.056 |
| | | Objects shared by actors working in same artistic genre | 0.1935*** | 0.068 | 0.4014*** | 0.08 |
| Socio-material | | Engagement with same materiality | 0.0275 | 0.039 | 0.0166 | 0.04 |
| | | Usage of objects that are part of material contexts | 1.5527** | 0.734 | 2.5003*** | 0.929 |
| | | Influence of social ties on usage of objects | 0.1378*** | 0.05 | -0.1845*** | 0.03 |
| | | Influence of social ties on usage of objects in dyads | -0.0173*** | 0.003 | - | - |
| | | *Influence of dyadic social ties on object sharing [H1]* | 0.9735* | 0.58 | 0.677 | 0.586 |
| | | *Influence of dyadic social ties between actors of same gender on objects sharing [H2]* | 0.0206 | 0.051 | -0.0985* | 0.053 |
| | | *Influence of dyadic social ties between actors of similar education on objects sharing [H2]* | 0.3512*** | 0.088 | 0.5563*** | 0.069 |
| | | *Influence of dyadic social ties between actors working in same artistic genre on objects sharing [H2]* | 0.1059** | 0.052 | -0.1951*** | 0.064 |
| | | *Influence of dyadic social ties on engagement with the same material context [H3]* | -0.1198*** | 0.039 | 0.3265*** | 0.076 |

**Note**: Black circles: actors. Cyan triangles: objects. Circles' numbering indicates attributes of actors.
Unstandardized coefficients; two-tailed tests reported; *p<0.1; **p<0.05; ***p<0.01

## 4.3. Structuring of social, material and object usage networks

Social networks exhibit some principles of structure formation well known in social network analysis. In both of the types of social networks, there are tendencies towards triadic clustering, with particular confidence for collaborations, which means that artists are likely to work in triads. Collaboration networks also tend to decentralize, as the negative parameter for *Degree distribution* indicates, which is to be expected in creative groups with distributed leadership. There is no gender homophily in the networks of artists. Genre homophily is found in the emotional attachment networks (artists tend to be friends with those working in the same genre), but not in the collaboration networks. Artists are known to need support and comments from each other and share information (Farrell, 2003), but do not tend to work with competitors who produce similar art as this would lead to repetition, generally not valued in creative settings. Such a logic seems to work even stronger with regard to education: There is a clear tendency against working with those who experienced similar training. Indeed, collaborating with those who were taught the same approaches and techniques has less potential to result in something new.

In the material networks of the collectives, similar principles are found. Consistently through both of the models, there is a strong tendency against centralization in the networks of objects and a strong tendency towards triadic closure in the material context networks. Shared objects filling the five artistic spaces tend to be combined in triads and there is no tendency for any objects to be more focal than the rest.

For the networks of object usage, several significant effects are found as well. Expectably, actors tend to use several objects and to combine the objects they use. Although pairs of actors do happen to share objects, actors refrain from focusing on some particular objects rather than the rest. In other words, material engagement tends to be spread more or less equally among the actors. Furthermore, sharing of objects is affected by all three attributes of the actors: individuals of the same gender, working in similar genres and having similar education all tend to share objects. In other words, similarity in characteristics leads to similarity of material choices.

While controlling for social and material structuring processes outlined above, the models also capture significant parameters for a number of more complex patterns that simultaneously include social ties, usages of objects and relationships between objects. Such patterns depict relationships between social structure and materiality. In the remainder of this section I focus on those corresponding to my hypotheses.

### 4.4. Influence of dyadic social ties on objects sharing (H1)

With regard to H1a, there is evidence that collaborations encourage sharing of objects (*Influence of dyadic social ties on object sharing:* 0.9735, p<0.1; note, however, that the parameter is marginally significant). Hence, *H1a is confirmed*.

The ethnographic data captures a variety of ways collaborators come to use the same artistic tools and materials in the shared spaces. For instance, consider collaborators MI and MD[4], members of the Madrid collective that mainly creates installations out of trash. These two members are most often found executing the construction part of the collective's projects together and sharing a lot of tools and materials in the process. For instance, in 2015 they worked on a project where children made a tangram (a kind of brain-teaser) out of cardboard boxes. They both used the tool for cutting cardboard into pieces so that they have a specific form needed to make the tangram. First, MI made measurements and cut wooden planks with a circular saw for the wooden part of the tool. Then, MD took some metal planks and cut them with the same saw in order to make the metal part of the tool. In the end, they made a press with a specific pattern to cut cardboard pieces for the tangram. Later the same day, they took turns working with the CC cutter - a computer-operated machine that cuts precise figures out of wood. They used it to cut pictures of faces, drawn by children of a school with which they collaborate. First, MD programmed the scanned images into the machine and fixed a wide wooden plank on the panel of the cutter and then MI controlled the process of cutting.

Whereas joint practice is a more obvious way to objects sharing, a similar outcome is achieved by collaborators in separate practices. In a parallel project of preparation for an exhibition, MI decided to use cardboard boxes for building a kind of a cave inside the hall of the gallery, so that the visitor felt as if inside a pile of old cardboard trash. MI used an electric drill to screw the boxes to a wooden carcass framing the entrance to the gallery (Figure 6). A few days later, we saw MD also use the drill to fix his boxes to planks for the tangram construction project. Perhaps, the idea to use the drill for this purpose came from observing MI at work, yet it is used for a different project in a different setting. Thus, collaborators mimic each other's usages of objects, producing sequences of adoptions. In this process, similar materiality is explicitly engaged in.

---

[4] Here and further, the names of individuals are encoded, the first letter in the code coming from the first letter in the name of the city where the collective is located and the second randomly assigned to the informants.

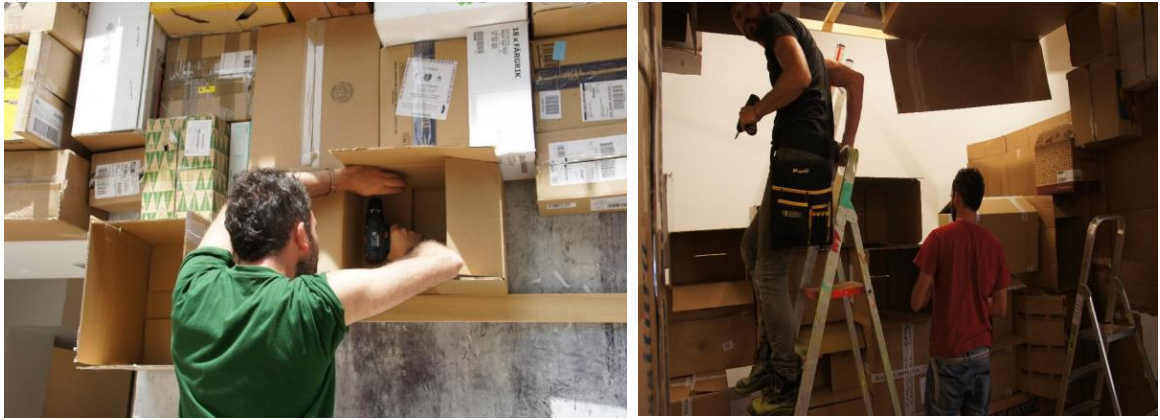

**FIGURE 6. MI AND MD USING CARDBOARD BOXES AND ELECTRIC DRILL IN THEIR PROJECTS**

**Note:** MI: Left hand side. MD: Right hand side.

Note that the two persons are well aware of the fact that they share tools. Shown a picture of the table with the heavy tools during the photo elicitation (Figure 7), MD comments:

> "This is, basically, the space where we keep our tools. I use it a lot for a lot of things. There are loads of kinds of working tools and I am one of those who use them the most. Officially, they are for everyone who works in the workshop, but actually only [MI] and myself use them."

This is also noticed by other members of the collective. For instance, MB in his photo elicitation comments on the same picture:

> "The tools for cutting wood and metal when it is necessary for a project. [MI] and [MD] are those who use them the most."

Note, however, that the members do not specify sharing of any particular tools and do not necessarily realize how they come to share them. It may very well be that the principle bringing them to share physical objects is not reflected by the collaborators themselves or the other members.

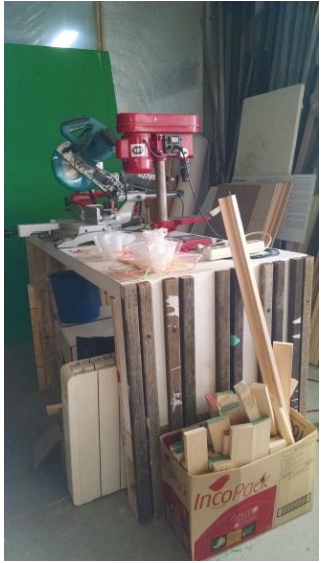

FIGURE 7. HEAVY TOOLS IN THE MADRID SPACE

Meanwhile, the Emotional attachments model does not provide evidence of a similar principle operative with regard to dyads of friends, who do not tend to share things (parameter for the *Influence of dyadic social ties on object sharing* pattern is not significant in the Emotional attachments model). Therefore, *H1b is rejected*.

Moreover, mere usage of objects is negatively impacted by the amount of friends individuals have (*Influence of social ties on usage of objects* in Emotional attachments model: -0.1845, p<0.01). This result suggests that friendships negatively affect overall individuals' engagement with materiality. At the same time, having more collaborators stimulates actors to use more items. These results suggest that engagement with the same materiality is stimulated by collaborative social ties, but not by friendships. However, further results reveal a more complex picture.

### 4.5. Influence of homophilous social ties on object sharing (H2)

Education homophily appears to mediate the relationship between social ties and object usage. It turns out that not only collaborators who received a similar education (and hence are trained to use certain techniques and approaches) are likely to use the same objects (*Influence of dyadic social ties between actors of similar education on objects sharing* in the Collaborations model: 0.3512, p<0.01), but friends of similar education are as well (0.5563, p<0.01). Thus, both *H2a and H2b are supported for education*.

This is despite the overall tendency for artists with a similar education to refrain from collaboration, striving for originality, and the absence of education homophily in friendship networks. Whereas plain education homophily has little effect on the formation of socio-material networks in the empirical setting under study, education homophilous ties, when they do occur, stimulate 'contagion with materiality'.

As to genre homophily, it rather duplicates the mechanisms described by the first pair of hypotheses. Similarity of artistic genres individuals primarily work in increases the likelihood of collaborators to pick the same objects in their common artistic spaces (*Influence of dyadic social ties between actors working in same artistic genre on objects sharing:* 0.1059, $p<0.05$ for the Collaboration model and -0.1951, $p<0.01$ for the Emotional attachments model). For instance, if two collaborators in the Hamburg collective, both doing photography, prepare an exhibition together, they are especially likely to share objects. Simultaneously, genre-homophilous friendships are associated with dissimilar object usage. Therefore, *with regard to genre H2a is supported, but H2b is rejected.*

In addition to the absence of gender homophily in the networks of emotional attachments, when individuals of the same gender do not tend to become friends, they also tend not to share objects (*Influence of dyadic social ties between actors of same gender on objects sharing* in the Emotional attachments mode: -0.0985, $p<0.1$; note, however, that the level of confidence is not very high). There is also no tendency towards object sharing among collaborators of the same gender (parameter for the respective pattern is not significant). Hence, both *H2a and H2b are rejected for gender.*

Accounting for competition between multiple processes taking place in socio-material networks also allows more light to be shed on the interplay of contagion with regard to object use with homophily, gender and career trajectory effects. For instance, the *Emotional attachments* model reveals a genre homophily effect in the friendship networks: Artists working in the same genre tend to be friends so that they can share advice and support each other in the career challenges common to them. At the same time, artists working in the same genre also tend to use similar objects, as they do similar things technically. Despite the latter tendency, those who engage in genre-homophilous relations do not tend to use the same objects. Although the general tendency for artists working in the same genre is to use the same things, engaging in affective relationships they do not resemble each other's material choices.

Concerning education, the relationships between homophily, contagion, and career trajectory effects in friendship contexts are different. The effect corresponding to education homophily is not significant, and no impact of training on usage of the same objects is observed. Similarity in training trajectories is thus insufficient to stimulate embeddedness in the same materiality. Meanwhile, education-homophilous friendships, when they do occur, are related to usage of the same objects. Hence, these ties are to do with socio-material contagion. In contrast to the emotional attachment networks, artists tend not to collaborate with those who received similar training. Yet still, when education-homophilous collaborations do occur, they are associated with usage of the same objects (note that this happens above the controlled tendency of actors of similar training to use the same items).

In sum, controlling for homophily, gender and career trajectory effects in social and object usage networks, it turns out that homophilous social ties affect dyadic engagement with common materiality.

Note that similarly to the results of the first pair of hypotheses, the present findings also suggest that, unlike collaborations, emotional attachments seem to be negatively associated with shared objects usage (surplus to attribute-based patterns, see *Influence of social ties on usage of objects* in the Emotional attachments model: -0.1845, p<0.01). From the perspective proposed in my conceptual section, this is a puzzling finding. However, bear in mind that collaborations of the visual artists often involve work with physical materials. Therefore, the more collaborative projects – the more shared objects are to be used by an actor. And when a dyad is collaborating, the members stimulate each other to use similar sets of tools or materials to create their joint artworks, organize exhibitions together, etc., especially when they have received similar training and when they work in the same genre. For instance, two collaborators who are professionally trained artists and do mosaics are likely to prefer a cutter to a brush and use the cutter to produce mosaic pieces, not installation elements. Moreover, they also engage with the same everyday items, such as dishes, furniture, household items and food when they work and take short breaks, having tea or a beer in between. Meanwhile, friendly relationships are rather realized as casual encounters, far less requiring physical things and drawing on a few everyday items that are often different – a couple of (separate) cups, a pair of (separate) chairs, etc. Does this further support the conclusion that friendships have little to do with material embeddedness?

During the ethnographic studies, a number of situations supporting this assumption were witnessed. Consider the material activities of two friends from the St. Petersburg group, SA and SF. The two have known each other for many years and have been friends since their childhood. Indeed, they often meet and chat. However, we continuously observed them engaging with different objects, even working on the same projects in the same spaces. For instance, SA and SF were in the same room for hours preparing a stage performance. However, the sets objects they used were different. SA drank instant coffee using an electric kettle at a bar counter, took food out of a cupboard, and discussed the performance with his collaborator SH. Meanwhile, right next to SA, SF created decorations for the performance in the same room, using veneer sheets, pencils, a measuring tape, a circular saw and other things.

In another situation, when the collective members were making coffins for a performance in pairs, SA and SF also worked using different sets of tools (Figure 8).

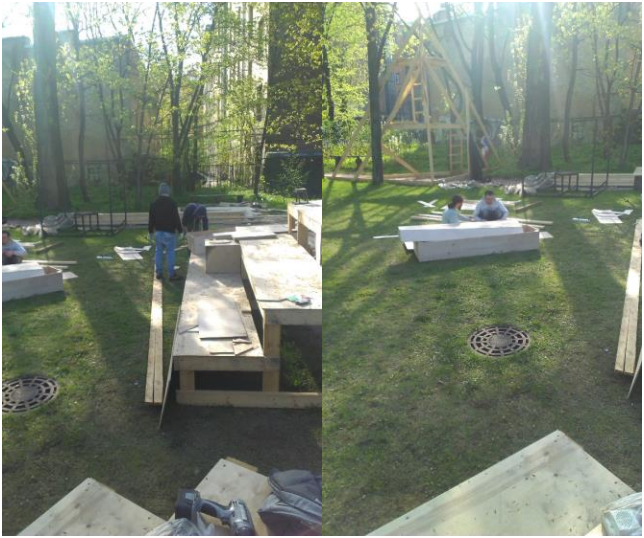

FIGURE 8. SA AND SF DURING PREPARATION OF THE COFFIN PERFORMANCE

**Note:** Left hand side: SA working on a coffin with his collaborator. Right hand side: SF working on a coffin with his collaborator.

A month later, during the preparation for an indoor performance, we observed SA and SF using visibly different sets of tools during the whole process. SF used tools that transform the exhibition space, such as a stapler, a stepladder and a glue gun. SA rather worked with electronic objects: set up a computer, handled a projector and selected videos to show on a TV during the performance.

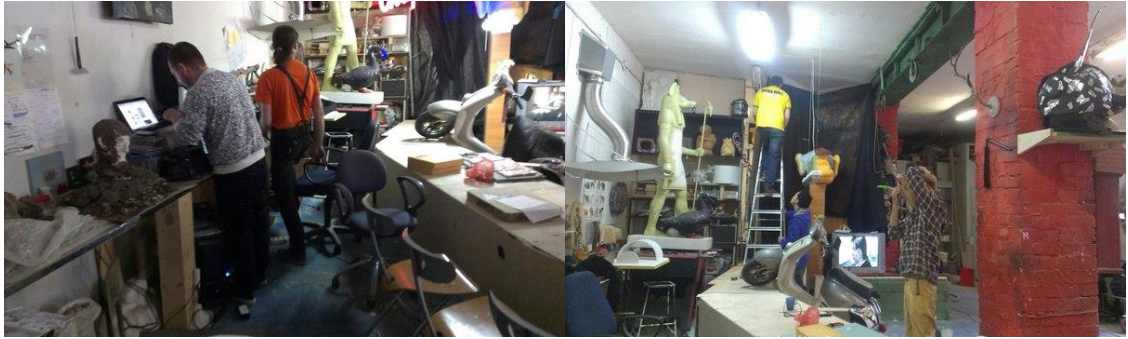

FIGURE 9. SA AND SF DURING PREPARATION OF THE INDOOR PERFORMANCE

**Note:** Left hand side: SA setting up the computer. Right hand side: SF and decorating the room on the stepladder.

In an observation of casual interaction, SA and SF were, again, in the same room for hours, but engaged with different objects, SA reconsidering his old artworks and SF working at his computer. When another member, SE, arrived, SA engaged in discussion of one of his works, whereas SF was playing around with his artwork (skateboard) instead.

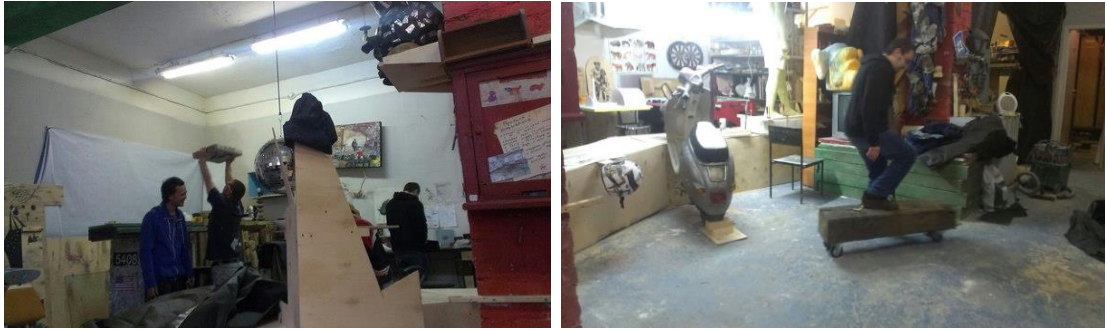

**FIGURE 10. SA AND SF DURING CASUAL INTERACTIONS**

**Note:** Left hand side: SA and SE discuss an artwork and SF is doing his work alone. Right hand side: SF rides his scooter as SA and SE chat

In sum, in contrast to my conceptual speculation, it seems that social ties do not necessarily have to do with common materiality, at least as far as friendships are considered. However, such a conclusion would not be correct taking into account the findings with regard to engagement of dyadic social ties with the same material contexts.

### *4.6. Influence of dyadic social ties on engagement with the same material context (H3)*

A positive and significant parameter for the pattern *Influence of dyadic social ties on engagement with the same material context* in the *Emotional attachments* model (0.3265, p<0.01) suggests quite the opposite: A tendency occurs for actors to engage with a particular area in the shared physical space the more their friends do. Simultaneously, the more actors' collaborators appear to engage with specific material contexts the less the actors engage with these contexts (*Influence of dyadic social ties on engagement with the same material context* in the *Collaborations* model: -0.1198, p<0.01). Hence, *H3a is rejected, whereas H3b is supported*. While mere selection of objects is stimulated by working relationships, engagement with common material contexts is connected with friendship.

The process observed in the Collaborations model can be seen in the light of two motivations typical for the creative settings: division of labor and striving of artists for novelty. First, collaborators tend to split tasks in their common projects and often even work on these tasks at different times. Relying on the same set of tools and resources, just like we see in the example from the Madrid collective, imitating each other and hence replicating each other in the items they use, they utilize them to accomplish different tasks – and hence combine these physical things differently in the course of their activities. An example are MI and MD from the Madrid collective when they share the same heavy tools to accomplish different aspects of a joint project or when they work on different projects using the same set of heavy tools. Second, while co-workers do use the same tools and materials, perhaps they do not combine the items with those also used by their collaborators because they try to make something differently using similar sets of common items at their disposal that they tend to share. So that while certain items are shared, the materiality they engage with distinguishes the artists. In two of the collectives under study (Barcelona and Hamburg) artists tend to create artworks individually, so the latter motivation is more relevant to them, whereas the former is more applicable to the other three of the collectives (London, Madrid and St. Petersburg), in which artworks are jointly created.

In contrast to collaborations, we know from the other findings that friends draw on objects individually – favorite cups and spoons, preferred drinks, personally comfortable chairs, and so on. However, these objects used solely by individuals tend to be part of the same material contexts. Although the work of this principle is more difficult to notice in the flow of everyday life and it is the totality of socio-material relations analyzed statistically that reveals them, based on the modelling results we know where to look.

Consider an example of BU and BK, a dyad of artists who soon after joining the Barcelona collective discovered many common interests and became friends. During three weeks of field observations field researchers traced both individual and joint material practices of BU and BK. It was found that during meetings in the kitchen, in the patio at lunch, at dinners, and during parties, they did not use the same objects, but continuously operated within the same material contexts, where the objects are interrelated.

For instance, at one of the artists' dinner gatherings, BU and BK were seen moving around, joining one conversation or another, but never sitting on the same chair or bench. BU preferred to stand or sit on the bench in the patio and BK chose to use one of the plastic patio chairs (Figure 11). During daily routine, field researchers observed how BU and BK used the shared kitchen (Figure 12). Sometimes they coincided there and had lunch together, but normally they did not. BU used to come to the shared studio space early in the morning and drink a lot of coffee, using the common cups available in the kitchen and the coffee machine. Normally, she took a coffee capsule, made a cup of coffee and went to her studio. BK prefers to drink tea from her own cup. So, when she decided to have a break and went to the kitchen, she took her own cup, put a teabag in it, boiled water in a kettle (that BU does not use but which has a connection with the common kitchen cups as these are located and often used together with the kettle by other members), poured water into her cup, and, if nothing interrupted her, went back to her studio.

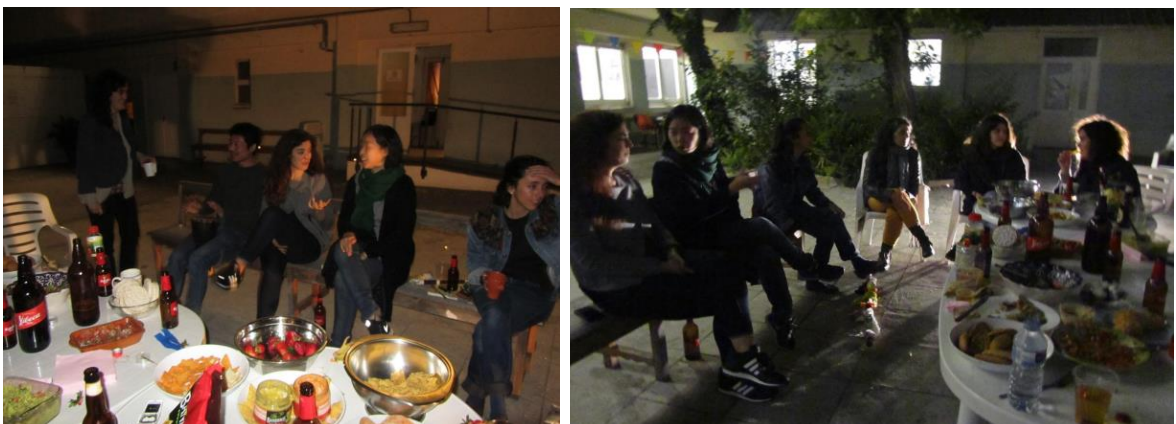

**FIGURE 11. BU AND BK DURING DINNER**

**Note:** Left hand side: BU. Right hand side: BK

Similar examples are found in the other collectives under study. For instance, in the example of SA and SF in St. Petersburg we also found that the different sets of objects used by the two friends are functionally related and normally placed together.

In sum, the ways joint material embeddedness of friends is achieved makes a contrast to that of collaborative relationships, which explicitly embody themselves in the physical world as shared usage of objects in joint work. Friends seem to rather follow different trajectories in their material practices and pinpoint certain objects as personally theirs, but in doing so they eventually embed in the same material context. Mere joint embeddedness in the same material space appears to be crucial for friendship, rather than the sharing of particular objects. Perhaps, affective relationships tend to extend themselves in the physical world indirectly, via links to objects that are related in the material context.

As the ethnographic data also show, the work of the socio-material structuring principle covered by *H3b* is almost invisible, hiding in the nuances of seemingly trivial everyday actions and banal relations between material objects, in which friends embed, without reflecting on it. During fieldwork and examining ethnographies, we can find how this principle is operative only when we know what we are looking for from the statistical analysis that captures the mechanisms behind the flow of everyday life.

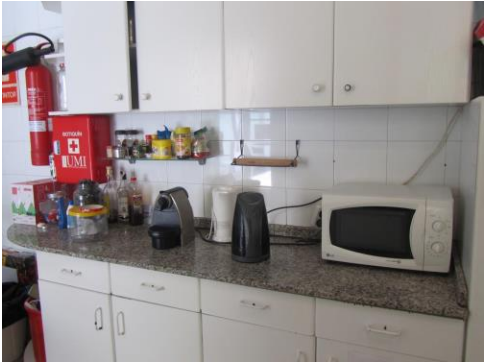

FIGURE 12. KITCHEN IN THE BARCELONA SPACE

These different activities yield different patterns of engagement with material objects that are part of persistent material relations (Figure 13).

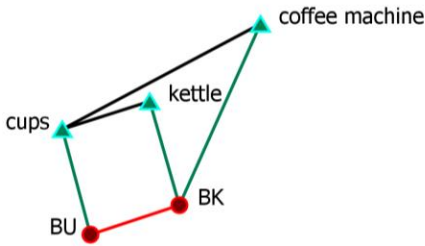

FIGURE 13. SOCIO-MATERIAL PATTERNS OF BU'S AND BK'S KITCHEN PRACTICES

**Note:** Circles: actors. Triangles: objects. Red line: friendship tie. Black lines: material links. Cyan lines: usages of objects.

# 5. Conclusion

The paper extended the duality perspective (Breiger, 1974; Breiger and Pattison, 1986; Breiger and Puetz, 2015; Fararo and Doreian, 1984; Martin, 2000; Mohr, 2000; Mohr and Duquenne, 1997; Mohr and Neely, 2009) to the relationship between the social and the material orders. I proposed a mixed method socio-material network analysis to examine the interplay between physical objects and social ties using data on five artistic collectives located in different European cities. Applying statistical network modeling to the relationship between social network structure and materiality, and using the two-mode approach enabled inferring specific principles of how interpersonal ties, both work-related and everyday, lead to shared materiality.

Although my hypotheses based on well-known principles, such as homophily and contagion, might seem commonsense, statistical network analysis of the ethnographic data showed that with regard to materiality the work of these principles is not so obvious. In particular, they may work or not work with regard to different types of social ties and different actor attributes. Namely, dyads of friends tend to implicitly embed in physical contexts via usage of different objects part of these contexts, as friends hang out and engage in everyday activities, but not sharing single items. Artists tend to be distinct from their friends in their choices of physical things. Yet, despite these tendencies, similar education trajectories may stimulate friends to use the same items. Meanwhile, working relationships discourage engagement in common material contexts, as collaborators split tasks between them and prefer practices that distinguish them from those they work with. However, usage of shared items is stimulated by joint work, with similar career trajectories reinforcing this relationship. In short, both collaborative and expressive relationships stimulate individuals to share materiality, but in different ways. Instrumental ties lead to sharing of specific items. Expressive ties stimulate embeddedness in shared material contexts.

These are only the first findings in the direction of socio-material network analysis and a number of limitations and future prospects can be identified. Indeed, the findings may be relevant only to creative settings, where aesthetically-sensitive individuals working in visual arts are particularly active in molding their material environment. Similarly, the results may be limited to small groups of collocated members.

Several methodological extensions are possible. First of all, one must take into account that ethnographic data on objects usage is inevitably culturally mediated because of the presence of observer and the reporting procedures involved. Hence, a degree of approximation remains as more robust techniques delivering data on object usage are yet to be found. Besides, further research is to consider ways to take into account the effect of culture on the interplay between social ties and materiality.

Furthermore, the strength of a mixed ethnographic dataset is that it contains not only reported material practices, but also the ones observed by field researchers. Hence, by contrast to purely report-based studies, it may cover many habitual practices overseen by informants. Often, usages of tools, materials, dishes, and furniture are not being recognized. Indeed, in our experience, informants do not report many of the links field researchers observed. A subsequent temptation is to compare the observed data to the reported ones in order to draw conclusions on how habitualized material practices relate to purposeful actions. However, I must admit that the current dataset does not allow for checking whether a particular instance of object usage is not recognized or whether it is merely not reported. Posing questions about absent links and comparing them to observed links could be one of the interesting paths to take this kind of analysis further.

In addition, two waves of data have been analyzed here. A longitudinal analysis examining the dual interplay between the social and the material is to follow. While here I only accounted for the impact of social structure on the material structure, there may also be an inverse impact of objects sharing and common practices on social ties between actors, as actor-network theory argues (Callon et al., 1986; Latour, 2005).

    Bearing all of the limitations in mind, the aim of this paper was to unveil the opportunities opened up by a mixed method network-analytical approach to the relation between social and material orders, rather than to draw conclusions. The combination of ethnographies and statistical modeling has a lot more to shed light on.

# Appendix A: Goodness of fit tests for MERGMs

I performed the test on the parameter values in the models plus a number of other patterns not parametrized in the models by calculating t-values. For simulations, after a 100,000 iteration burn-in, I picked a sample of 10,000 networks from 100,000 iteration simulations produced using the fitted models. For parameters included in the models t-ratios up to 0.1 indicate a good model fit, for parameters not included in the models–t-ratios below 1.0. As the tests indicate, the models do not completely explain clustering in the B network (the network of objects).

|  | **Collaborations** | | | | **Emotional attachments** | | | |
|---|---|---|---|---|---|---|---|---|
| *Statistics* | *Observed* | *Mean* | *SD* | *t-ratio* | *Observed* | *Mean* | *SD* | *t-ratio* |
| EdgeA | 64 | 62.4 | 11.204 | 0.143 | 78 | 79.1303 | 13.219 | -0.086 |
| Star2A | 214 | 214.74 | 77.42 | -0.01 | 384 | 400.4455 | 179.433 | -0.092 |
| Star3A | 222 | 255.3 | 148.291 | -0.225 | 732 | 790.6369 | 625.495 | -0.094 |
| Star4A | 155 | 230.668 | 204.847 | -0.369 | 1111 | 1246.252 | 1385.403 | -0.098 |
| Star5A | 70 | 166.811 | 223.363 | -0.433 | 1316 | 1541.21 | 2125.66 | -0.106 |
| TriangleA | 47 | 49.261 | 17.307 | -0.131 | 81 | 89.0266 | 63.016 | -0.127 |
| Cycle4A | 76 | 107.184 | 61.432 | -0.508 | 301 | 390.8658 | 475.045 | -0.189 |
| IsolatesA | 4 | 2.744 | 2.017 | 0.623 | 2 | 2.2251 | 1.416 | -0.159 |
| IsolateEdgesA | 0 | 1.203 | 1.165 | -1.032 | 0 | 0.4837 | 0.699 | -0.692 |
| ASA | 134.0625 | 128.8884 | 35.916 | 0.144 | 185.6797 | 190.1148 | 49.959 | -0.089 |
| ASA2 | 134.0625 | 128.8884 | 35.916 | 0.144 | 185.6797 | 190.1148 | 49.959 | -0.089 |
| ATA | 91.5 | 88.3135 | 23.765 | 0.134 | 115.5469 | 118.2822 | 31.626 | -0.086 |
| A2PA | 150.5 | 135.6274 | 40.665 | 0.366 | 198.2344 | 198.9471 | 22.475 | -0.032 |
| AETA | 214.8125 | 236.2452 | 100.393 | -0.213 | 438.5898 | 486.4391 | 382.26 | -0.125 |
| Gender_MatchA | 37 | 36.261 | 7.05 | 0.105 | 41 | 41.5238 | 6.979 | -0.075 |
| Gender_MismatchA | 27 | 26.139 | 6.068 | 0.142 | 37 | 37.6065 | 7.485 | -0.081 |
| Education_MatchA | 54 | 52.396 | 10.651 | 0.151 | 75 | 76.0894 | 13.105 | -0.083 |
| Education_MismatchA | 10 | 10.004 | 2.104 | -0.002 | 3 | 3.0409 | 1.795 | -0.023 |
| Genre_MatchA | 20 | 19.794 | 3.673 | 0.056 | 20 | 20.0531 | 2.744 | -0.019 |
| Genre_MismatchA | 44 | 42.606 | 9.361 | 0.149 | 58 | 59.0772 | 12.444 | -0.087 |
| EdgeB | 52 | 50.763 | 13.055 | 0.095 | 52 | 51.5452 | 10.838 | 0.042 |
| Star2B | 101 | 100.105 | 80.036 | 0.011 | 101 | 99.2158 | 61.641 | 0.029 |
| Star3B | 75 | 100.747 | 177.086 | -0.145 | 75 | 87.5119 | 194.645 | -0.064 |
| Star4B | 32 | 109.219 | 326.169 | -0.237 | 32 | 94.6095 | 886.154 | -0.071 |
| Star5B | 6 | 118.808 | 510.024 | -0.221 | 6 | 175.4582 | 4012.817 | -0.042 |
| TriangleB | 26 | 17.815 | 12.501 | 0.655 | 26 | 17.5846 | 9.09 | 0.926 |
| Cycle4B | 48 | 21.226 | 36.349 | 0.737 | 48 | 16.5718 | 22.452 | 1.4 |
| IsolatesB | 42 | 43.481 | 6.087 | -0.243 | 42 | 44.0408 | 5.581 | -0.366 |
| IsolatedEdgesB | 7 | 8.313 | 2.789 | -0.471 | 7 | 7.2377 | 2.512 | -0.095 |
| ASB | 70.75 | 67.2821 | 36.832 | 0.094 | 70.75 | 69.3411 | 28.906 | 0.049 |
| ASB2 | 70.75 | 67.2821 | 36.832 | 0.094 | 70.75 | 69.3411 | 28.906 | 0.049 |
| ATB | 43.125 | 41.2227 | 22.194 | 0.086 | 43.125 | 42.4008 | 17.577 | 0.041 |
| A2PB | 66.125 | 82.1697 | 55.158 | -0.291 | 66.125 | 84.4591 | 45.934 | -0.399 |
| AETB | 119.5 | 65.3764 | 68.663 | 0.788 | 119.5 | 60.695 | 47.529 | 1.237 |
| XEdge | 511 | 510.002 | 35.819 | 0.028 | 511 | 516.2066 | 52.026 | -0.1 |
| XStar2A | 4637 | 4617.346 | 827.058 | 0.024 | 4637 | 4898.531 | 1375.978 | -0.19 |
| XStar2B | 1611 | 1558.411 | 216.783 | 0.243 | 1611 | 1627.647 | 448.074 | -0.037 |
| XStar3A | 33805 | 32943.56 | 10317.88 | 0.083 | 33805 | 37842.75 | 18713.14 | -0.216 |
| XStar3B | 3768 | 3466.376 | 702.936 | 0.429 | 3768 | 3821.496 | 1869.552 | -0.029 |
| X3Path | 63956 | 64351.78 | 16697.16 | -0.024 | 63956 | 72732.55 | 37258.33 | -0.236 |
| X4Cycle | 11466 | 10598.14 | 3494.884 | 0.248 | 11466 | 12760.04 | 9382.781 | -0.138 |
| XECA | 512444 | 481377.9 | 230335.1 | 0.135 | 512444 | 664829.4 | 665396.4 | -0.229 |
| XECB | 168426 | 152367.3 | 60612.11 | 0.265 | 168426 | 206289.2 | 205409.7 | -0.184 |
| IsolatesXA | 1 | 0.872 | 0.865 | 0.148 | 1 | 1.0906 | 0.995 | -0.091 |
| IsolatesXB | 0 | 0.159 | 0.395 | -0.403 | 0 | 0.2979 | 0.626 | -0.476 |
| XASA | 872.9583 | 871.122 | 70.827 | 0.026 | 872.9583 | 883.3801 | 103.493 | -0.101 |
| XASB | 679.605 | 677.3433 | 67.748 | 0.033 | 679.605 | 689.8479 | 102.493 | -0.1 |
| XACA | 1922.0852 | 1925.665 | 139.085 | -0.026 | 1922.085 | 1964.511 | 84.963 | -0.499 |
| XACB | 291.542 | 279.2415 | 13.007 | 0.946 | 291.542 | 291.3349 | 8.443 | 0.025 |
| XAECA | 45764.3603 | 42340.09 | 13979.2 | 0.245 | 45764.36 | 50973.48 | 37531.09 | -0.139 |
| XAECB | 43899.1934 | 40942.99 | 13995.98 | 0.211 | 43899.19 | 49621.03 | 37588.78 | -0.152 |
| Gender_X2StarAMatch | 852 | 845.829 | 115.521 | 0.053 | 852 | 873.3529 | 212.733 | -0.1 |
| Gender_X2StarAMismatch | 759 | 712.582 | 107.824 | 0.43 | 759 | 754.2942 | 237.392 | 0.02 |
| Gender_X4CycleAMatch | 5787 | 5918.129 | 1941.024 | -0.068 | 5787 | 6790.278 | 4549.974 | -0.221 |
| Gender_X4CycleAMismatch | 5679 | 4680.01 | 1632.303 | 0.612 | 5679 | 5969.761 | 4852.13 | -0.06 |
| Education_X2StarAMatch | 1529 | 1514.132 | 215.492 | 0.069 | 1529 | 1574.482 | 447.465 | -0.102 |
| Education_X2StarAMismatch | 82 | 44.279 | 11.466 | 3.29 | 82 | 53.1651 | 12.624 | 2.284 |
| Education_X4CycleAMatch | 11270 | 10535.27 | 3493.418 | 0.21 | 11270 | 12671.18 | 9381.961 | -0.149 |
| Education_X4CycleAMismatch | 196 | 62.871 | 28.829 | 4.618 | 196 | 88.8544 | 37.922 | 2.825 |
| Genre_X2StarAMatch | 330 | 328.309 | 44.061 | 0.038 | 330 | 336.9611 | 67.847 | -0.103 |
| Genre_X2StarAMismatch | 1281 | 1230.102 | 177.863 | 0.286 | 1281 | 1290.686 | 383.104 | -0.025 |
| Genre_X4CycleAMatch | 2190 | 2381.608 | 774.565 | -0.247 | 2190 | 2751.11 | 1508.807 | -0.372 |
| Genre_X4CycleAMismatch | 9276 | 8216.531 | 2762.996 | 0.383 | 9276 | 10008.93 | 7894.747 | -0.093 |
| Star2AX | 1752 | 1683.879 | 418.212 | 0.163 | 2858 | 2965.892 | 1055.773 | -0.102 |
| StarAA1X | 1955.9063 | 1790.458 | 673.904 | 0.246 | 3882.967 | 4063.169 | 1894.206 | -0.095 |
| StarAX1A | 3003.5159 | 2876.875 | 751.936 | 0.168 | 5102.552 | 5307.168 | 2013.425 | -0.102 |
| StarAXAA | 1125.1181 | 1117.687 | 76.532 | 0.097 | 1181.221 | 1196.754 | 149.811 | -0.104 |
| TriangleXAX | 601 | 555.902 | 151.895 | 0.297 | 980 | 1028.907 | 540.512 | -0.09 |
| L3XAX | 14820 | 14050.14 | 4345.278 | 0.177 | 30200 | 35069.66 | 21038.8 | -0.231 |
| ATXAX | 120.6689 | 117.5432 | 22.781 | 0.137 | 149.3956 | 151.7254 | 26.298 | -0.089 |
| EXTA | 2048 | 2059.394 | 947.325 | -0.012 | 5659 | 6661.801 | 6726.775 | -0.149 |
| Gender_MatchTXAX | 347 | 334.189 | 92.119 | 0.139 | 485 | 510.1052 | 252.213 | -0.1 |
| Gender_MismatchTXAX | 254 | 221.713 | 78.165 | 0.413 | 495 | 518.8015 | 294.336 | -0.081 |
| Education_MatchTXAX | 548 | 523.671 | 150.689 | 0.161 | 966 | 1021.31 | 540.475 | -0.102 |
| Education_MismatchTXAX | 53 | 32.231 | 10.933 | 1.9 | 14 | 7.5964 | 5.779 | 1.108 |
| Genre_MatchTXAX | 179 | 175.099 | 53.198 | 0.073 | 206 | 211.1687 | 61.268 | -0.084 |
| Genre_MismatchTXAX | 422 | 380.803 | 121.785 | 0.338 | 774 | 817.738 | 489.833 | -0.089 |
| Star2BX | 781 | 753.604 | 249.152 | 0.11 | 781 | 783.0195 | 216.702 | -0.009 |
| StarAB1X | 622.75 | 567.0112 | 359.441 | 0.155 | 622.75 | 589.1754 | 282.575 | 0.119 |
| StarAX1B | 1162.897 | 1113.552 | 398.527 | 0.124 | 1162.897 | 1165.067 | 359.988 | -0.006 |
| StarAXAB | 880.4806 | 875.1319 | 95.351 | 0.056 | 880.4806 | 891.3593 | 118.99 | -0.091 |
| TriangleXBX | 337 | 265.286 | 99.404 | 0.721 | 337 | 267.0193 | 100.704 | 0.695 |
| L3XBX | 3302 | 2992.745 | 1246.417 | 0.248 | 3302 | 3112.833 | 1313.929 | 0.144 |
| ATXBX | 95.2939 | 92.8239 | 25.472 | 0.097 | 95.2939 | 94.59 | 20.822 | 0.034 |
| EXTB | 766 | 455.044 | 366.321 | 0.849 | 766 | 442.5212 | 262.476 | 1.232 |
| L3AXB | 2570 | 2690.399 | 1449.674 | -0.083 | 4768 | 5088.482 | 2604.642 | -0.123 |
| C4AXB | 911 | 917.873 | 534.461 | -0.013 | 1826 | 1765.752 | 1247.001 | 0.048 |
| ASAXASB | 2578.6563 | 2357.469 | 892.476 | 0.248 | 4505.717 | 4652.344 | 1998.802 | -0.073 |
| AC4AXB | 1212.334 | 1192.166 | 121.848 | 0.166 | 1434.58 | 1437.466 | 56.54 | -0.051 |
| stddev_degreeA | 2.518 | 2.5144 | 0.358 | 0.01 | 3.4641 | 3.4199 | 0.718 | 0.061 |
| skew_degreeA | 1.395 | 1.5418 | 0.146 | -1.003 | 1.7691 | 1.6393 | 0.136 | 0.951 |
| clusteringA | 0.6589 | 0.693 | 0.063 | -0.541 | 0.6328 | 0.6165 | 0.116 | 0.141 |
| stddev_degreeX_A | 9.0169 | 8.8709 | 1.172 | 0.125 | 9.0169 | 9.3305 | 1.647 | -0.19 |
| skew_degreeX_A | 0.7691 | 0.5394 | 0.146 | 1.571 | 0.7691 | 0.684 | 0.118 | 0.72 |
| stddev_degreeX_B | 3.2047 | 3.006 | 0.25 | 0.796 | 3.2047 | 3.0792 | 0.531 | 0.237 |
| skew_degreeX_B | 0.6812 | 0.6125 | 0.214 | 0.321 | 0.6812 | 0.6067 | 0.182 | 0.408 |
| clusteringX | 0.7171 | 0.6486 | 0.045 | 1.537 | 0.7171 | 0.6546 | 0.094 | 0.666 |
| stddev_degreeB | 1.537 | 1.46 | 0.412 | 0.187 | 1.537 | 1.4806 | 0.313 | 0.18 |
| skew_degreeB | 2.1439 | 2.0937 | 0.328 | 0.153 | 2.1439 | 2.0877 | 0.307 | 0.183 |
| clusteringB | 0.7723 | 0.5555 | 0.098 | 2.206 | 0.7723 | 0.5466 | 0.089 | 2.546 |

# Appendix B: Correlations between the estimates of insignificant independent variables

| **Collaborations model** | Edge | Ties between actors of same gender | Ties between actors working in same genre |
|---|---|---|---|
| Edge | | | |
| Ties between actors of same gender | -0.2445 | | |
| Ties between actors working in same genre | 0.2014 | 0.0193 | |

| **Emotional attachments model** | Degree distribution | Ties between actors of same gender | Ties between actors of similar education |
|---|---|---|---|
| Degree distribution | | | |
| Ties between actors of same gender | 0.0479 | | |
| Ties between actors of similar education | -0.0385 | -0.1646 | |